# A connected and automated vehicle readiness framework to support road authorities for C-ITS services

Bahman Madadi[a]*

Ary P. Silvano[b]

Kevin McPherson[c]

John McCarthy[d]

Risto Öörni[e]

Gonçalo Homem de Almeida Correia[a]

[a] Delft University of Technology, Delft, The Netherlands

[b] Swedish National Road and Transport Research Institute, Linköping, Sweden

[c] Transport Research Laboratory (TRL), London, United Kingdom

[c] ARUP, Dublin, Republic of Ireland

[c] Technical Research Centre of Finland (VTT), Espoo, Finland

* Corresponding author: b.madadi@tudelft.nl

2023

## ABSTRACT

Connected and Automated Vehicles (CAVs) can have a profound influence on transport systems. However, most levels of automation and connectivity require support from the road infrastructure. Additional support such as Cooperative Intelligent Transport Systems (C-ITS) services can facilitate safe and efficient traffic, and alleviate the environmental impacts of road surface vehicles. However, due to the rapidly evolving technology, C-ITS service deployment requirements are not always clear. Furthermore, the costs and benefits of infrastructure investments are subject to tremendous uncertainty. This study articulates the requirements using a structured approach to propose a CAV-Readiness Framework (CRF). The main purpose of the CRF is allowing road authorities to assess their physical and digital infrastructure readiness, define requirements for C-ITS services, and identify future development paths to reach higher levels of readiness to support CAVs by enabling C-ITS services. The CRF is intended to guide and support road authorities' investment decisions on infrastructure.

# 1 Introduction

Whilst connected and automated driving (CAD) is expected to have a profound influence on transport systems in terms of road safety, traffic efficiency, and environmental impacts, it is an area of technology that is likely to cause disruption and change in the design and operation of road network infrastructure as well. Most forms of CAD require some level of infrastructure support for their safe operation [1]. Additional infrastructure and services to support CAD would have the potential to improve safety even further and to bring other benefits such as reduced congestion [2], [3]. National road authorities (NRAs) can influence the deployment of CAVs along with the extent and the direction of their impacts by providing suitable physical and digital infrastructure as well as harnessing the power of CAD data to improve traffic safety, efficiency, and environmental impacts [4], [5].

Various infrastructure-based solutions for enhancing CAD have been suggested in the literature. For instance, the concept of Infrastructure-Enabled Autonomy (IEA) has been suggested in [6], which proposes enhanced automated driving via roadside infrastructure. V2X communications for infrastructure-assisted CAD was suggested in [7]. Infrastructure Support levels for Automated Driving (ISAD), which was developed within the INFRAMIX project [8], Automated Drivability indeX (ADX) proposed in [9], and physical and digital infrastructure readiness index for CAD [4] are among other concepts that can aid road authorities gauge their infrastructure's readiness to support CAD. Other studies have provided extensive lists of possible infrastructure adjustments to support CAD based on literature reviews [1], [10]–[13] and expert opinions [14], [15].

However, making extensive infrastructure upgrades can be idealistic, costly, and unnecessary in some places, particularly in rural areas and local urban streets whose main functionality is providing accessibility to origins and destinations [1], [16]. In addition, conventional road infrastructure is already costly, for example, the total infrastructure costs for road, rail and waterway transport in the EU28 countries amount to approximately €267 billion in 2016 [17]. Accommodating the needs of new CAD technologies and enabling different Cooperative Intelligent Transport Systems (C-ITS) services may increase these costs even further. Moreover, given the rapidly evolving CAD technology and uncertain projections of future CAD demand, the infrastructure requirements for supporting CAD and C-ITS services are not always clear, and it is difficult for NRAs to predict and plan the future levels of support needed. To make informed investment decisions, NRAs need to be able to accurately assess requirements, costs, and benefits of C-ITS services.

Recognizing the cost-benefit trade-off of infrastructure investment decisions for CAD, many studies have attempted to utilise optimization or cost-benefit analysis to determine the most cost-effective network-wide plan of action for the deployment of optimal networks of dedicated CAV lanes [18], [19], dedicated CAV links [20], [21], dedicated CAV zones [22], CAV-ready links for mixed traffic [16], [23], and mixed configurations [24]–[27]. However, these studies mostly focus on connected highway and urban driving scenarios and do not consider other C-ITS use cases and services, such as Hazardous Location Notifications (HLN), Green Light Optimal Speed Advisory (GLOSA), Road Works Warning (RWW), and probe vehicle data.

Moreover, interviews conducted for this research with European NRAs revealed that not all NRAs are interested in all CAD scenarios and C-ITS services. Based on the specific geographical, economic, socio-political, and cultural characteristics of their regions, different NRAs have unique needs, requirements, and aspirations. Therefore, guiding infrastructure investments of different NRAs requires an understanding of their needs through engagements and developing flexible frameworks tailored to their specific situation.

Therefore, this study proposes a flexible CAV-readiness framework (CRF) that enables NRAs to have a broad overview of all possible C-ITS services they can provide to support CAD, the requirements of each service, and their potential costs and benefits. The CRF recognises heterogeneous needs and aspirations of NRAs in different regions and influences the ability of the NRAs to become digital road operators. It can guide them not only to plan infrastructure projects but also to develop a long-term strategy for their networks.

The structure of this manuscript is as follows. The next section elaborately describes the CRF. Section 3 includes a case study to demonstrate how CRF could be used in practice. The closing section includes the concluding remarks and recommendations.

## 2    CAV-readiness framework (CRF)

The CRF we propose and develop is intended to be used by NRAs to help assess their readiness to support CAD, and to implement specific C-ITS services and use cases. The CRF uses a hierarchical structure (shown in Figure 1) to breakdown services and use cases into their smallest building blocks, i.e., enablers, which are the lowest building blocks of the CRF tool and could be seen as infrastructure elements. In short, CRF includes the following steps (visualised in Figure 2):

- Identifying the C-ITS services to be provided and the implementation scenarios based on the objectives of the NRA;
- Breaking those services down into smaller building blocks such as use cases and enablers;
- Scoring the NRAs readiness, aspirations based on their specific needs and goals, and high-level assessment of costs and contributions of each enabler to help plan and prioritise the NRA support for CAD;
- Providing useful information regarding use case costs and impacts as well as the NRA readiness at various levels.

In this section, we elaborate on each step individually. Figure 1 and Figure 2 provide visual summaries of the CRF structure and steps involved with its implementation.

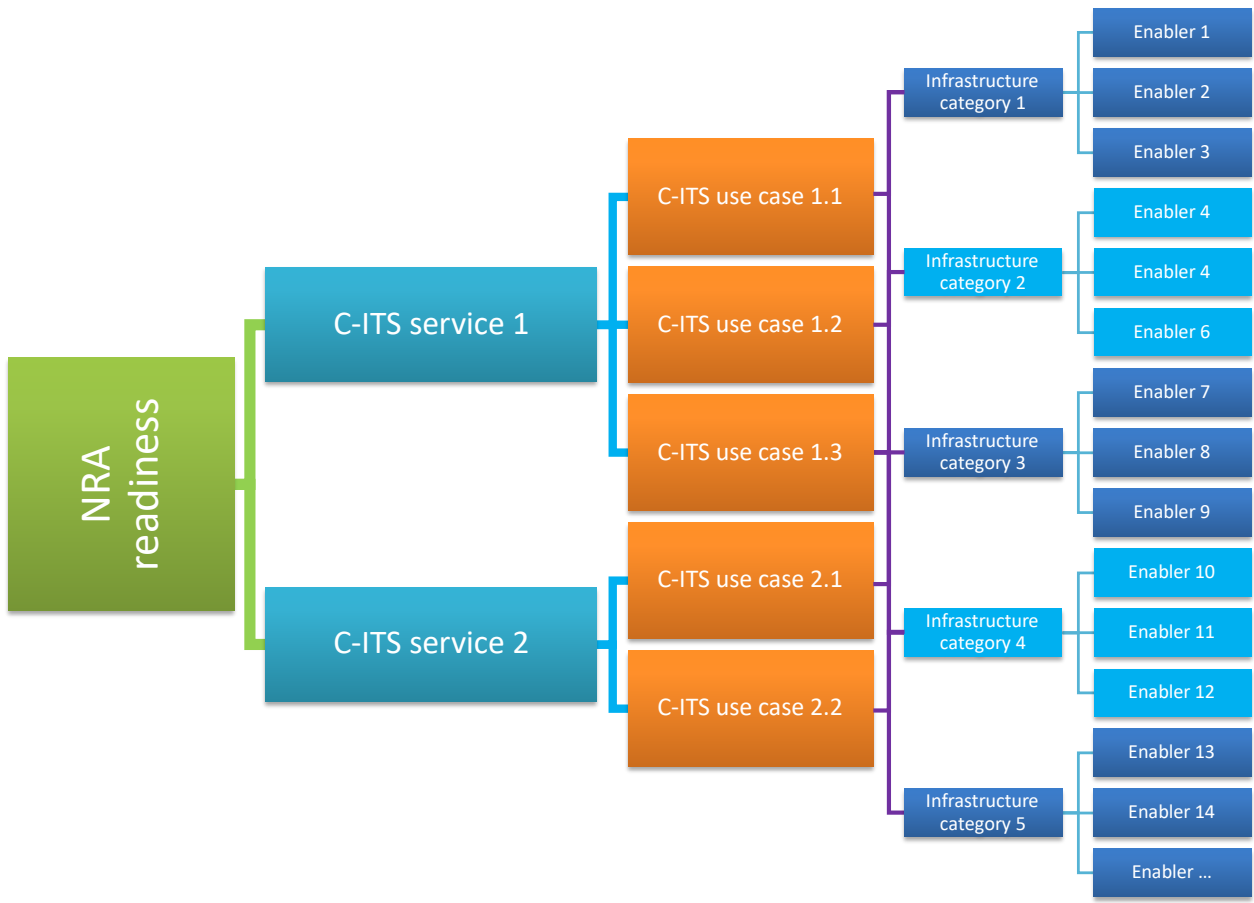

*Figure 1  The CRF structure*

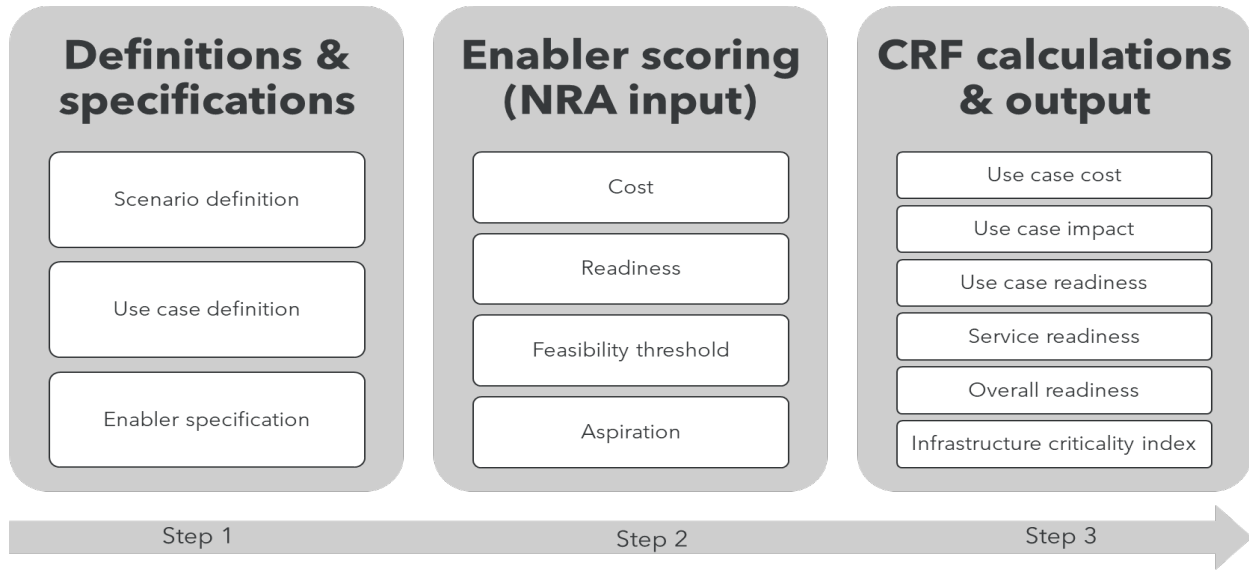

*Figure 2 Summary of CRF steps*

## 2.1 Use case definition

The CRF is structured on C-ITS services and use cases as defined under the C-ROADS definitions [28]. C-ROADS is a joint initiative of European Member States and road operators for testing and implementing C-ITS services, with a desire for cross-border harmonisation and interoperability. In C-ITS terminology, a service is a clustering of use cases based on a common denominator, for example, an objective such as "awareness of road works". Services are also known as applications. Under C-ROADS, the deployment of C-ITS is seen as evolutionary, starting with less complex services ("Day-1 Services" encompassing messages about traffic jams, hazardous locations, road works, and slow or stationary vehicles, as well as weather information and speed advice). "Day 2" and "Day 3+" services are being investigated in R&D projects. Hence the C-ITS Service and Use Case definitions [28] provide a firm basis on which to implement the CRF, both now and in the future.

Table 1 lists the main C-ITS services according to C-ROADS. Table 2 lists the use cases under RWW service. There is a one-to-many relationship between services and use cases. Take as an example the C-ITS Road Work Warning (RWW) service. C-ROADS currently identifies four use cases within that service: lane closure (RWW-LC), road closure (RWW-RC), road works – mobile (RWW-RM) and winter maintenance (RWW-WM).

Not all use cases are important for all NRAs. For instance, winter maintenance warnings may not be a concern in places with mild or no winter. Therefore, each specific NRA might be interested in a subset of services and use cases. This is recognized by the CRF that we propose by allowing for heterogeneous NRA aspirations through the possibility to select a subset of services and use cases of interest. They can then use the CRF to assess their readiness and draw an implementation roadmap tailored to their specific needs and requirements.

*Table 1 Main C-ITS services according to C-ROADS*

| C-ITS Services |
|---|
| In-Vehicle Signage (IVS) |
| Hazardous Location Notification (HLN) |
| Road Works Warning (RWW) |
| Signalized Intersections (SI) |
| Automated Vehicle Guidance (AVG) |
| Probe Vehicle Data (PVD) |

*Table 2   Use Cases for the RWW C-ITS Service*

| C-ITS Service | Use Cases |
|---|---|
| Road Work Warnings (RWW) | Lane closure (RWW - LC) |
| | Road closure (RWW - RC) |
| | Road works - Mobile (RWW - RM) |
| | Winter maintenance (RWW - WM) |

## 2.2    Scenario definition

Different implementation scenarios are possible for each C-ITS use case and each NRA might prefer one or more implementation scenario(s) to others. For instance, NRAs might decide to aggregate and archive C-ITS messages only on a national access point (NAP) or regional or privately managed clouds as well. Also, specific infrastructure elements can be provided by private stakeholders (e.g., OEMs) or the public. Moreover, each NRA might decide to implement a specific selection of use cases as a bundle, rather than all use cases in a service, based on their unique priorities, goals and aspiration. Each scenario will have different requirements. The CRF supports NRAs by allowing them to define their implementation scenarios and assess their readiness based on scenarios of interest.

## 2.3    Enabler specification based on C-ITS message information flow

For the purposes of the CRF, we systematically break down and describe each C-ITS use case using a set of enablers. The 'enabler' is the lowest building block of the CRF tool and could be seen as an infrastructure element. Some examples of enablers are shown in Table 3 (next section) for a demonstrative example along with their category, grouping them under one of following: Physical, Operation, Digital, Connectivity, or Standard. Although NRAs may not be the main owners of some specific enablers in some categories (e.g., standards), including the mentioned categories in CRF can benefit NRAs by providing them with an overview of the state of practice for enablers. An elaborate discussion regarding this topic is provided in the Discussion section. Figure 1 depicts the bottom-up structure of the CRF in terms of its building blocks. A detailed example is provided in the next section.

After defining enablers, we use a procedure based on the information exchange flow within each use case to specify enablers for each C-ITS use case. In short, the procedure traces C-ITS messages from the triggering event to the end user of the message, which is often the vehicle. For instance, in a simple case where an unplanned lane closure occurs due to an accident on the road, the information travels the path depicted in Figure 3 and listed next:

(1) The roadside unit (RSU) detects the accident and the lane closure due to it.

(2) The RSU sends a standardised message to the closest traffic control centre (TCC).

(3) The TCC determines relevant RSUs upstream of the accident to be informed.

(4) The TCC sends standardised messages to relevant RSUs.

(5) The relevant upstream RSUs informs vehicles nearby with standardised messages.

Each one of the above-mentioned steps requires certain enablers. For instance, step 1 requires RSUs that are equipped with sensors, algorithms, and processing power to detect accidents, a predefined list of events (e.g., lane closure) to look for, standard message profiles, e.g., Decentralized Environmental Notification Message (DENM) [29], and the connectivity infrastructure to transmit messages. Figure 3 depicts this example and points to relevant enablers for each step. Through using this procedure, relevant enablers for each use case and implementation scenario can be conveniently defined for any use case.

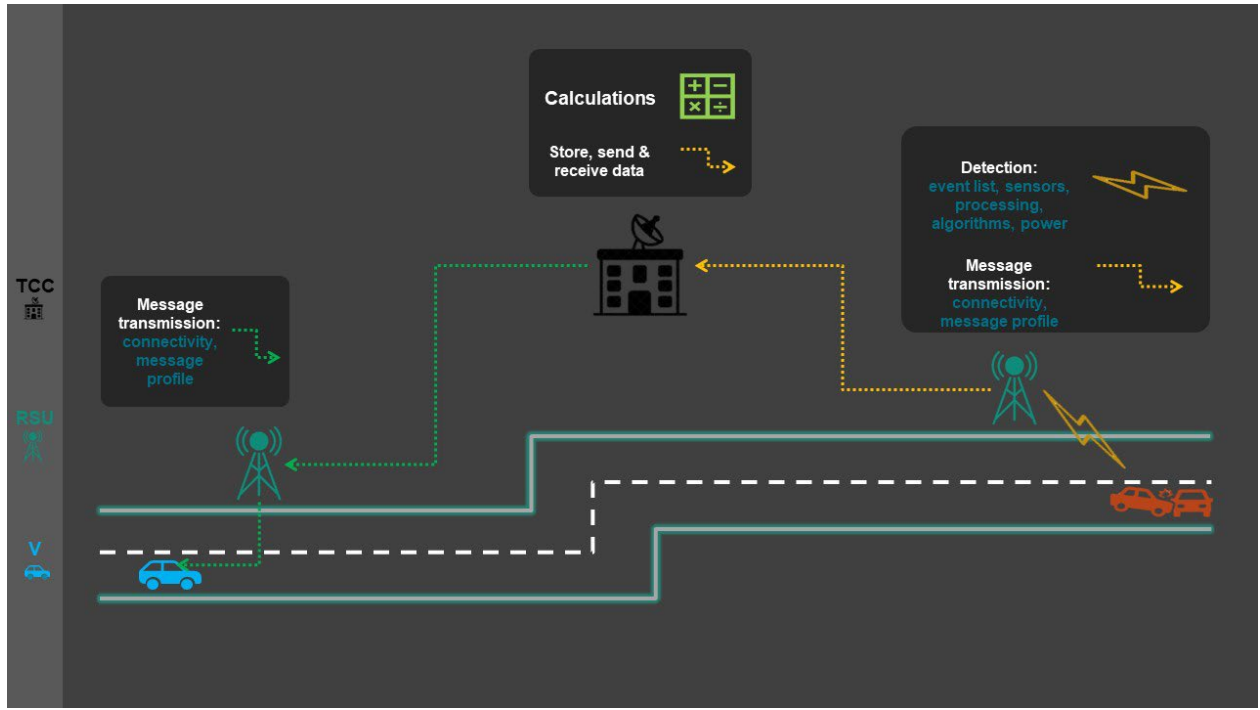

*Figure 3 Specifying enablers based on the information flow of C-ITS messages; in this example the information takes the following path through roadside units (RSU), the traffic control centre (TCC), and a vehicle (V): **RSU → TCC → RSU → V.***

## 2.4 Enabler scoring

For each enabler, the CRF should have a definition of:

- the readiness of the NRA to provide or deploy the enabler individually;
- the aspiration of the NRA to provide or deploy the enabler;
- the feasibility threshold for the service which defines the minimum level of support provided by the NRA to make implementation of this enabler feasible;
- high-level estimation of the costs and impacts of each enabler.

For scoring, we deploy a Level of Service (LoS) approach, which is a widely employed metric that quantifies the performance and quality of a provided service, utilizing a predetermined scale. The LoS definition is dependent upon the specific context and facility under examination, such as urban areas or motorways. These different environments may impose unique demands due to the traffic they accommodate. However, our goal in the CRF was to develop a generic LoS tool that could be applied to any road environment, to any use case or technology. The definition should therefore be generic, applicable to all cases, and technology agnostic. For C-ITS services (i.e., information provision), the aim is to provide information that is, among other things, accurate and timely to CAVs so they can react according to events on the road network. The CRF aims to illustrate the progress of (NRAs) towards becoming a digital authority, meaning that the NRAs should provide information (data provision) to CAVs that are precise, accurate, and timely.

However, many C-ITS services have not been implemented in practice long enough to have accurate estimations of their impacts as well as the cost and quality of their enablers. Therefore, we resort to a qualitative Likert-based approach to define three distinct levels of LoS and thereby the scores within the CRF. We first obtain scores within the values of {none, low, medium, high}. We define the importance of each enabler within the same scoring system as well. Then we transform this qualitative scheme to a quantitative one by first, mapping the values {none, low, medium, high} to the scores of {0, 1, 2, 3}, respectively, and then, multiplying each score by its importance to obtain a value in the range of [0, 9]. It should be noted that considering a score for importance is necessary since although all enablers are required for each use case, they do not have the same contribution and criticality for each use case. For instance, in the example described above, RSUs are clearly the most important enablers. Therefore, this should be recognised by the framework. The scoring system will be elaborated upon with an example in the next section.

The scores, costs, and impacts of each of the above can be rolled up to the level of the use case, the service, and indeed the total package of support for CAD, to help plan and prioritise NRA support for each enabler.

## 2.5   CRF output

Once the input is ready for the CRF, it can be fed to the framework to generate the desired output. Given the rather simple architecture and calculations required, different CRF implementations are possible in practice, for instance as a dashboard tool to be used by NRAs or a module within a digital twin. The main output components produced by the CRF are the following.

- Use case deployment cost: this is calculated by aggregating the cost score of all enablers required for deployment of the use case.

- Use case impact: this is simply a multidimensional illustration of different impacts (e.g., safety and environmental impacts) in one figure using the same scale for comparison purposes.

- Use case readiness: this is calculated based on the aggregation of all enablers for the use case, first, at the level of the category, then at the level of the use case. Similar scores are calculated for NRA aspiration and the feasibility threshold based on the aggregation of enablers to the category and then the use case level.

- Service readiness (progress): this simply shows the total progress (readiness) for each C-ITS service assuming each use case under the service has an equal contribution to the overall progress of the service.

- Overall readiness: this is calculated based on the aggregation of each category score for all use cases being considered by the NRA. A similar score is calculated for the overall aspiration to provide a measure of how close an NRA is to their aspired level of support for C-ITS services.

A visual summary of the CRF steps and structure is provided in Figure 4. In the following section, we present an illustrative example that guides the reader through the calculations and results obtained by the CRF.

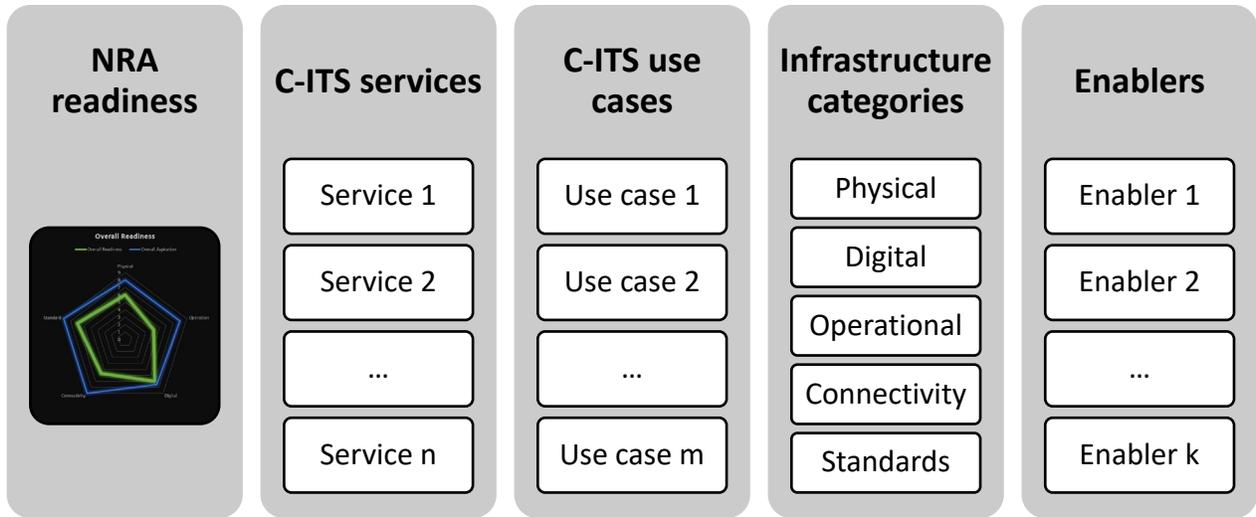

*Figure 4 A visual summary of CRF steps and structure*

# 3 Demonstration of CRF with a Road Works Warning (RWW) case study

In this section, we provide as an illustrative example, the design and implementation of an individual C-ITS service. This example is simplified in the way in which high-level enablers have been defined, in order to explain and visualise the workings of the framework. Figure 5 is derived from a Nordic Way Evaluation Report [30] based on an actual pilot project and describes the flow of messages from an implementation of a Road Works Warning (RWW) system. The service and warning message for RWW was generated at the RWW unit mounted on a truck mounted attenuator (TMA) vehicle. The message was received by a roadside unit which transferred the RWW message in DENM and DATEX II format [31] through an interchange node to the OEM cloud and then to the vehicle. The Original Equipment Manufacturers (OEMs) clouds also receive roadwork information messages from the Traffic Authority in DATEX II format.

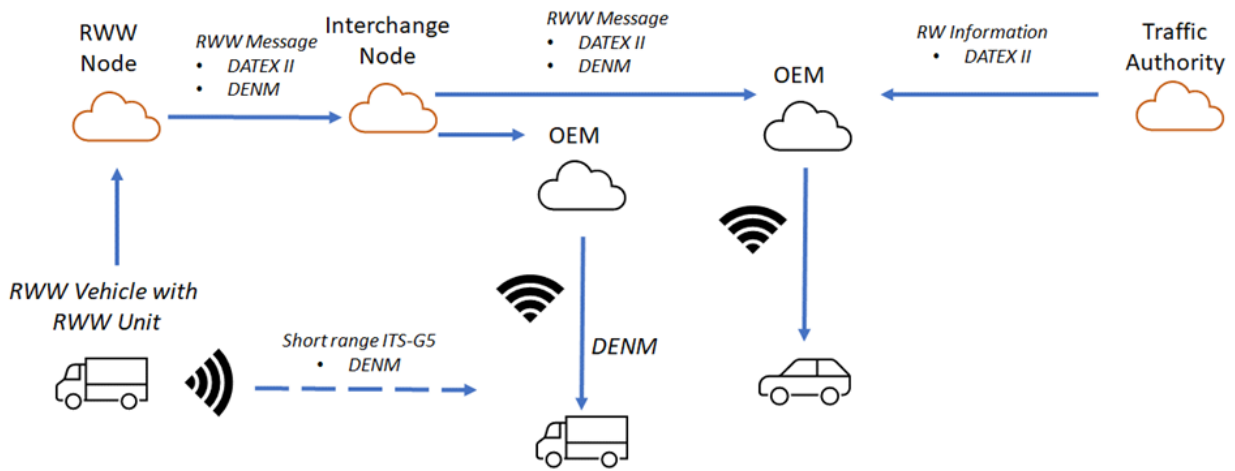

*Figure 5 Flow of messages for Road Works Warning (RWW) implementation based on the Nordic way 2 example*

Following the CRF steps described in the previous section, first, the use case and the implementation scenario are defined above (in this case, based on the actual pilot project implementation). The next step is using the C-ITS message information life-cycle procedure to define the enablers required for this use case. This procedure led to nine main enablers, which are listed in Table 3 along with their category.

*Table 3  Main enablers for Road Works Warning (RWW) implementation based on the Nordic way example*

| Enabler | Description | Category |
|---|---|---|
| ETSI EN 302 637-3 for DENM messaging | Standard for interoperability V2X LTE | Standard |
| ETSI TS 102 894-2 | Standard for sharing | Standard |
| Stationary RSU (R-ITS-S) | Stationary roadside unit | Physical |
| Mobile RSU (V-ITS-S) | TMA vehicle | Physical |
| Response plan | NRA procedure for triggering information exchange | Operation |
| R-ITS-S System Profile | System profile | Digital |
| V-ITS-S system profile | System profile | Digital |
| Cellular connectivity | Information from RSU to cloud | Connectivity |
| Short-range ITS-G5 | Information from TMA to RSU | Connectivity |

The next step is enabler scoring, which specifies the NRA readiness, aspiration, and feasibility threshold for deployment of each enabler for the purpose of this specific use case. In this case, we have assumed state of practice values just to demonstrate how CRF works and is used. Each enabler is considered as a separate item and is given an importance weighting (Low, Medium, High) within its use case. The readiness score is calculated as the multiplication of the importance of the enabler with the stated readiness of the NRA to provide or to deploy it (also Low / Medium / High) to give a Readiness Score. In addition to the readiness score, the framework also adds the concepts of the aspiration of the road authority to provide or deploy each enabler, and the feasibility threshold for the service which defines the minimum level of support provided by the NRA to make implementation of this use case feasible. The Aspiration and Feasibility Thresholds are given Low / Medium / High scores. A Feasibility calculation in the CRF helps to identify the enablers that the NRA needs to concentrate on to provide or deploy this use case. Table 4 and Table 5 summarise the information used for scoring, and the category scores obtained, respectively.

Within the CRF the average readiness, aspiration, and feasibility threshold of all enablers under the use case are shown diagrammatically (see Figure 6). The 'spokes' of the radar diagram represent the categories of the enablers.

The framework also allows the NRA to define the impact of deploying each use case, in terms of five key impact factors (cost, safety, efficiency, environment, and inclusion). Each of these impact factors is defined in terms of Low (1) / Medium (2) / High (3) scores. This graphically illustrates the relative costs and benefits of each use case or each service and can be used by the NRA to help prioritise the development or implementation of services (see Figure 7). Based on all the use cases that allows a specific NRA aspires to implement, overall readiness of the NRA can be defined per enabler category based on the average of the readiness and aspiration scores per category for all considered use cases. In our illustrative example, we have assumed that the NRA is only interested in one use case, which leads to the same overall readiness and aspiration (compare Figure 6 and Figure 8).

*Table 4  Enabler scoring for Road Works Warning (RWW) implementation based on the Nordic way example*

| Enabler | Category | Importance | Readiness | Readiness score | Aspiration | Aspiration score | Feasibility threshold | Threshold score | Cost |
|---|---|---|---|---|---|---|---|---|---|
| ETSI EN 302 637-3 for DENM messaging | Standard | High | High | 9 | High | 9 | Medium | 6 | None |
| ETSI TS 102 894-2 | Standard | High | High | 9 | High | 9 | Low | 3 | None |
| Stationary RSU (R-ITS-S) | Physical | High | Medium | 6 | High | 9 | Low | 3 | Medium |
| Mobile RSU (V-ITS-S) | Physical | High | Medium | 6 | High | 9 | Low | 3 | High |
| Response plan | Operation | High | Low | 3 | High | 9 | Low | 3 | Low |
| R-ITS-S System Profile | Digital | High | Medium | 6 | High | 9 | Medium | 6 | Low |
| V-ITS-S system profile | Digital | High | Low | 3 | High | 9 | Low | 3 | Low |
| Cellular connectivity | Connectivity | High | High | 9 | High | 9 | Low | 3 | Low |
| Short-range ITS-G5 | Connectivity | Medium | Medium | 4 | High | 6 | Low | 2 | Medium |

*Table 5  Infrastructure category scores for Road Works Warning (RWW) implementation based on the Nordic way example*

| Category | Readiness | Aspiration | Feasibility threshold |
|---|---|---|---|
| Physical | 6 | 9 | 3 |
| Operation | 3 | 9 | 3 |
| Digital | 4.5 | 9 | 4.5 |
| Connectivity | 6.5 | 7.5 | 2.5 |
| Standard | 9 | 9 | 4.5 |
| Total | 5.8 | 8.7 | 3.5 |

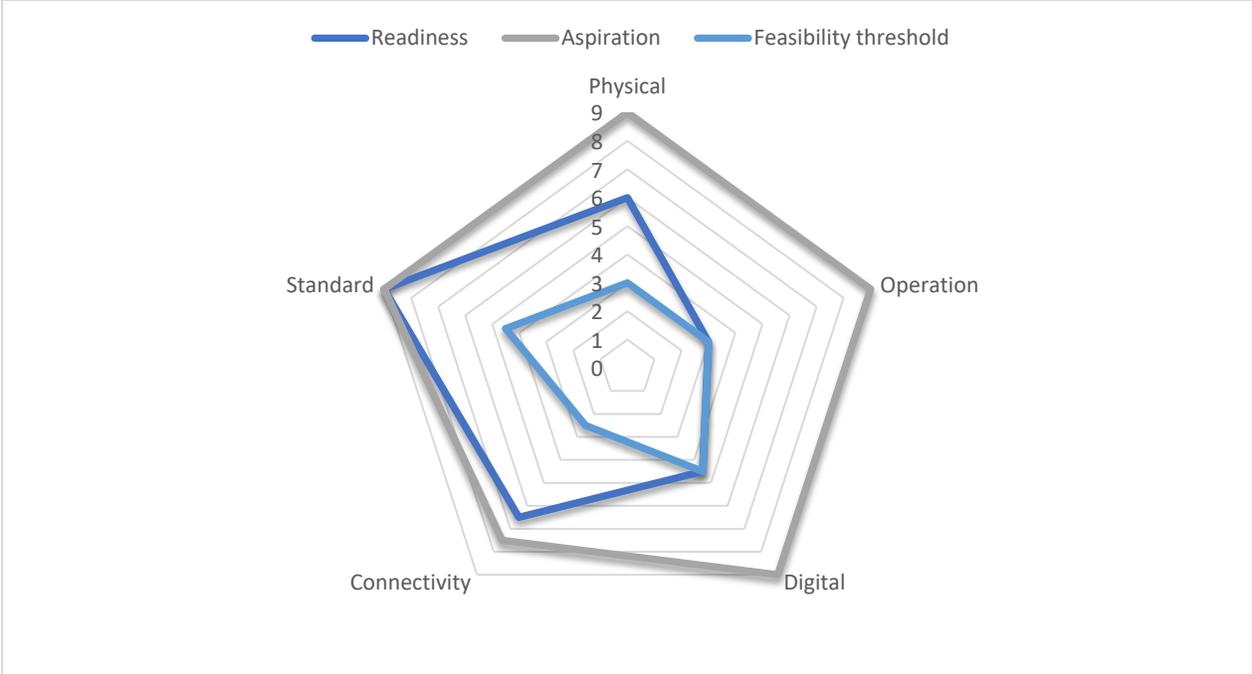

*Figure 6 NRA readiness for Road Works Warning (RWW) implementation based on the Nordic way example*

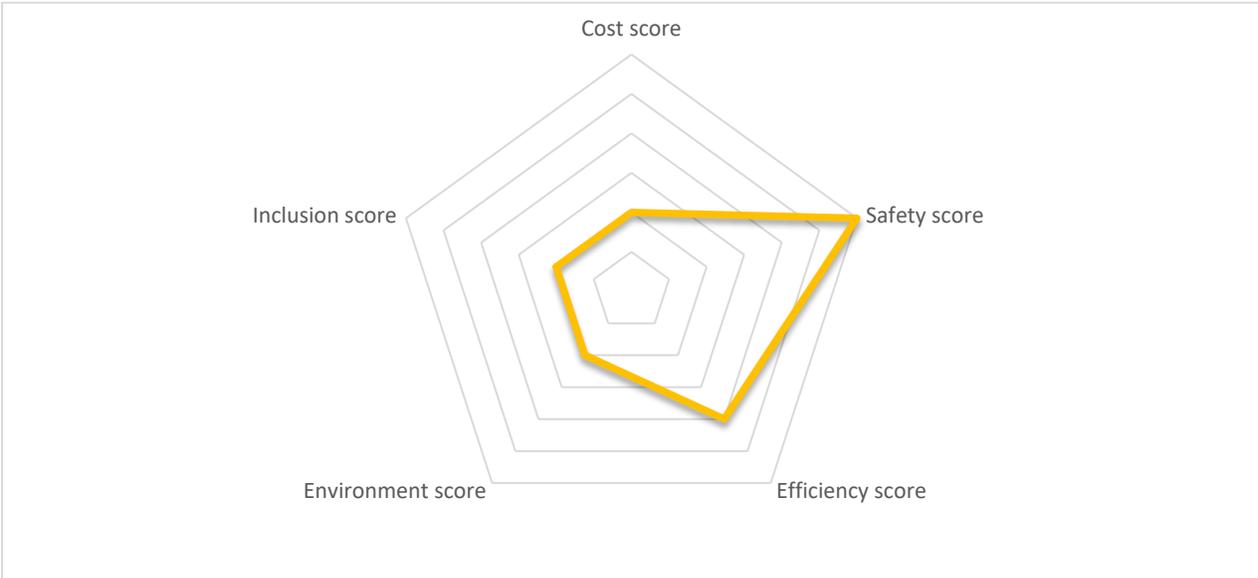

*Figure 7 Impacts of Road Works Warning (RWW) implementation based on the Nordic way example*

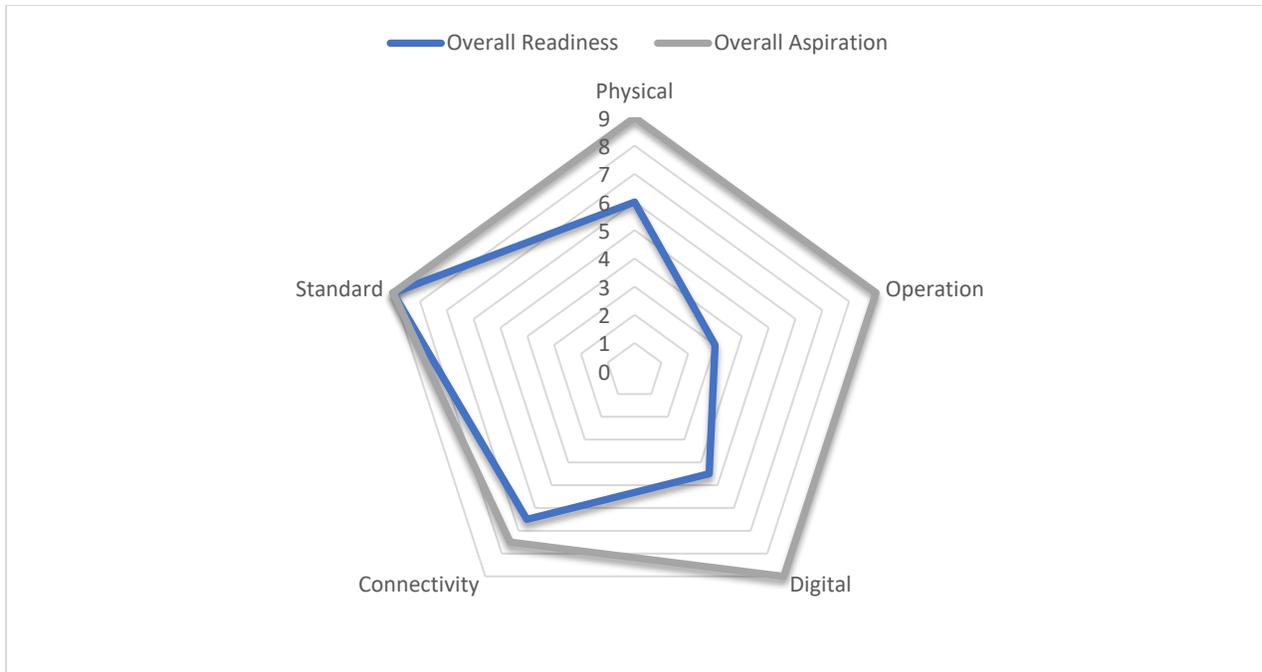

*Figure 8  NRA overall C-ITS readiness based only on the Road Works Warning (RWW) implementation from the Nordic way 2 example*

Expanding this concept up, hence scoring each of the use cases for the service, provides an assessment of an NRA's readiness to implement the entire C-ITS service. Should the NRA wish to support the deployment of the entire RWW service across its network, then the outputs of the CRF help it identify those use cases and enablers that it should prioritise. Figure 9 shows an illustrative example (not based on real data) of what could be the RWW service progress depiction based on assumed values for four use cases within this service. This provides NRAs with an overview of the progress of their use case and service deployment and aid them in prioritising their activities.

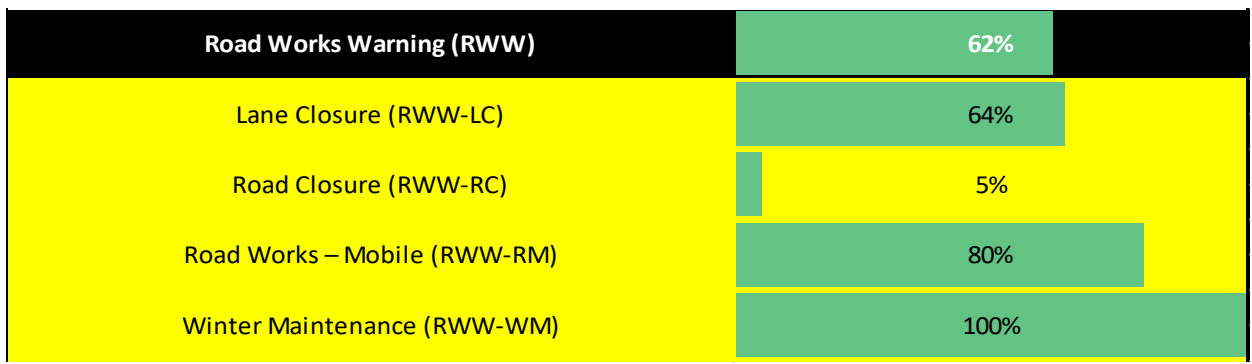

*Figure 9  RWW service progress illustration*

# 4    Discussion, conclusions, and future research directions

In this study, we introduced a flexible CAV-readiness framework (CRF) that empowers NRAs to gain a comprehensive understanding of all potential C-ITS services they can offer to bolster CAD, the prerequisites for each service, and their associated expenses and advantages. The CRF acknowledges the diverse requirements and ambitions of NRAs in various regions and impacts their capacity to evolve into digital road operators. It assists them in not only devising infrastructure investments but also linking the investments to an overall strategic approach to deployment and delivery of a range of services. The CRF decomposes each service into its fundamental building blocks, i.e., the enablers, and employs a structured bottom-up approach to evaluate the NRA's overall readiness. It also assesses the readiness of services and use cases across different infrastructure categories based on the enablers that constitute the use cases. The CRF encourages NRA engagement in the process by acquiring their current readiness and aspirations, and in that sense, it is tailored to the specific needs, requirements, and aspirations of each NRA.

Overall, the CRF can provide a good overview of where NRAs are, where they want to be, and specific areas to focus on to reach their specific goals. Moreover, it can aid in prioritizing NRA investments in infrastructure for CAD and C-ITS services. We showed its inner workings and application with one illustrative example, yet the CRF can do more. For example, the CRF could be used to compare the same use cases representing a basic level of service and an advanced level of service, using different enablers with different costs and benefits.

On the other hand, some factors influence NRAs' capacity to support CAD but are not considered by the CRF, thus it could be refined and improved in the future. The CRF is intended to be general and high-level to provide NRAs with an overview of where they are, and where they want to be, and how to get there in terms of a general roadmap. However, this means the costs and benefits of the implementation of each use case are not specific enough or sensitive enough to accommodate different levels of service. Also, cross-scenario comparisons lack the level of granularity required for formal cost-benefit analyses. In general, since most CAD technologies are not fully field tested yet, they have uncertain costs and benefits. Once NRAs start adopting the framework, different building blocks (e.g., costs estimated for specific enablers or impacts of specific use cases) can be refined and updated as more information becomes available through new projects and studies. The flexible nature of the CRF allows for evolution and refinement over time.

Another topic worth discussing is the enablers that are not necessarily in NRAs' control, e.g., standards and connectivity infrastructure. Consider as an example, standards, which are crucial enablers for implementing C-ITS services and without them, most services cannot be deployed. However, developing new standards is not necessarily under the control of NRAs. Therefore, the inclusion of this category in the CRF may not seem logical at first glance. Yet if for instance, the use of CRF makes it apparent that a standard is a bottleneck for the deployment of a C-ITS service, NRAs can become aware of it and push for its definition, despite not having direct control over it knowing that the use case depends on it. This is how including enabler categories that are governed by external stakeholders rather than NRAs themselves, such as the standards category, in the CRF can be useful for NRAs simply by providing them with an overview of which enablers are ready and which ones require more progress.

Lastly, the most important future work should be a large-scale and long-term application of CRF as a dashboard tool within an NRA digital twin to include all possible C-ITS services. The CRF must be gradually refined, both in terms of input and output, as new C-ITS service pilot projects and implementations provide more information and insight. Indeed, it is possible to use the CRF as the centrepiece of the debate to stimulate engagement and outcomes linked to the various questions raised when considering a roadmap for CAD deployment.

## Acknowledgement

The research presented in this paper is part of the DiREC project (consortium partners: TRL, ARUP, TU Delft, VTT, VTI and FEHRL) which is sponsored by the Conference for European Directors of Roads (CEDR) call for projects from 2020.